\documentclass[journal]{IEEEtran}
\usepackage{amsmath,amsfonts}

\usepackage{amssymb}
\usepackage{xcolor}

\usepackage{algorithmic}
\usepackage{algorithm}
\usepackage{array}
\usepackage[caption=false,font=normalsize,labelfont=sf,textfont=sf]{subfig}
\usepackage{textcomp}
\usepackage{stfloats}
\usepackage{url}
\usepackage{verbatim}
\usepackage{graphicx}
\usepackage{cite}
\usepackage{caption}
\def\BibTeX{{\rm B\kern-.05em{\sc i\kern-.025em b}\kern-.08em
    T\kern-.1667em\lower.7ex\hbox{E}\kern-.125emX}}
\usepackage{balance}
\begin{document}

\title{Robust UAV Jittering and Task Scheduling in Mobile Edge Computing With Data Compression}

\author{Bin Li, Xiao Zhu, and Junyi Wang
	\thanks{Bin Li and Xiao Zhu are with the School of Computer Science, Nanjing University of Information Science and Technology, Nanjing 210044, China (e-mail: bin.li@nuist.edu.cn).}
  \thanks{ Junyi Wang is with the School of Information and Communication, Guilin University of Electronic Technology, Guilin 541004, China (e-mail:
  wangjy@guet.edu.cn).}
  }

\markboth{xxx}%
{Shell \MakeLowercase{\textit{et al.}}: A Sample Article Using IEEEtran.cls for IEEE Journals}


\maketitle

\begin{abstract}
Data compression technology is able to reduce data size, which can be applied to lower the cost of task offloading in mobile edge computing (MEC). This paper addresses the practical challenges for robust trajectory and scheduling optimization based on data compression in the unmanned aerial vehicle (UAV)-assisted MEC, aiming to minimize the sum energy cost of terminal users while maintaining robust performance during UAV flight. Considering the non-convexity of the problem and the dynamic nature of the scenario, the optimization problem is reformulated as a Markov decision process. Then, a randomized ensembled double Q-learning (REDQ) algorithm is adopted to solve the issue.
The algorithm allows for higher feasible update-to-data ratio, enabling more effective learning from observed data. The simulation results show that the proposed scheme effectively reduces the energy consumption while ensuring flight robustness. Compared to the PPO and A2C algorithms, energy consumption is reduced by approximately $21.9\%$ and $35.4\%$, respectively. This method demonstrates significant advantages in complex environments and holds great potential for practical applications.

\end{abstract}


\section{Introduction}
Mobile edge computing (MEC) provides a faster and more efficient data processing method
for computation-intensive industrial applications by deploying computation servers near the end of devices.
However, it is inevitable that remote areas and rescue scenarios may bring high infrastructure deployment costs.
Integrating computing capabilities into unmanned aerial vehicle (UAV)  is increasingly 
capturing, attributed to the quick deployment and low cost \cite{UAVIntro}. 
As a mobile platform, the UAV can dynamically adjust their position and properly manage resource, 
significantly enhancing system flexibility and scalability.

In UAV-assisted MEC systems, effectively planning flight trajectories and optimizing task scheduling are crucial to ensuring the effective operation of the system. Particularly in energy-constrained environments, minimizing energy consumption becomes the core objective.
By designing optimal trajectories and scheduling strategies, energy consumption during flight can be reduced while improving the response time of computation tasks.

There are many exceptional works devoted to study the trajectory and scheduling optimization 
problems in UAV-assisted networks.
In \cite{UAVIntroE1}, the authors developed a novel optimization framework to optimize UAV 
trajectory, devices association and transmit power allocation with the aim of minimizing the 
energy consumption of all devices. 
In \cite{UAVIntroE2}, the authors achieved minimization 
of energy consumption and completion time of the UAV by jointly optimizing computation 
offloading, resource allocation, as well as UAV trajectory under the constraints of the 
devices' task and energy budget. 
The authors in \cite{UAVIntroE3} studied the joint optimization problem of task 
offloading, UAV trajectory, and resource allocation
while taking into account the interdependencies between different tasks,  with the aim of minimizing system energy consumption.
The authors in \cite{UAVIntroE4} studied the problem of priority-aware task offloading in a multi-UAV cooperation scenario, 
and formulated a joint optimization problem for UAV trajectory design, binary offloading decision-making, 
computing resource allocation, and communication resource management to maximize long-term average system gain.
Although the above works have effectively reduced the cost related to energy in the
MEC system, the battery capacities of the UAV and terminal users are still limited. Thus, 
how to further reduce the system energy consumption are crucial \cite{DRL2}.

Data compression technology has been widely applied to various types of files, such as text, audio, and video, 
not only saving storage resources but also reducing the burden of data transmission. 
In MEC systems, data compression has been explored as an effective means to reduce data size, 
thereby decreasing the energy and delay costs associated with task offloading \cite{compressionintro}. 
Building on this, several studies have examined the use of data compression techniques in MEC environments, 
highlighting their potential to improve system efficiency and performance.

In \cite{losslessCompress}, the authors adopted lossless compression technology and achieved the energy consumption minimization in multi-user MEC system by jointly optimizing computation offloading, data compression, and resource allocation.
In \cite{lossyCompress}, the authors studied joint offloading and compression decisions with 
the objective of optimizing the trade-off between the delay and accuracy of service requests 
based on deep learning services in the 3-tier user-edge-cloud system. To enhance the data 
processing capabilities of sensors, the authors in \cite{lossylossless} proposed a scheme 
for sensor-cloud systems based on edge computing, which realizes the lossy and lossless 
mixed compression of data. In \cite{Compress3Loc}, the authors studied and compared three 
different computation models with the aim of minimizing the weighted-sum delay of all devices.
In order to save the energy consumption and reduce the latency for wireless transmission, 
the authors in \cite{CompressRatio} investigated a multi-user MEC system using data compression 
technology to reduce the redundancy of sensing data. 
The above works adopt data compression technology to effectively reduce the cost of MEC system, but only considering the situation that edge server is deployed on the ground, which is not suitable for areas with complex terrain.
As such, the authors of \cite{UAVCompress} further studied the UAV-assisted MEC system and achieved the minimization of total energy consumption 
by jointly optimizing transmission power, task compression ratio, communication resource allocation, 
and UAV trajectory, while considering constraints such as task deadline and resource budgets. 
In this study, data compression technology played a key role in reducing energy consumption.

In UAV-assisted communication, the received signal power and data transmission rate are significantly affected by channel path loss, 
which is closely related to the air-to-ground link distance. The mobility of UAVs allows control of their position, 
thereby significantly improving the overall performance of the communication system. However, most studies in UAV trajectory optimization assume ideal physical conditions and overlook the potential jitter issues that UAV may experience in real-world environments.
In practical applications, some factors such as instability in the flight control system, variations in wind speed, 
and sensor errors may cause the UAV to deviate from its planned trajectory, thereby affecting its positioning accuracy and communication quality, which in turn reduces the stability and rate of data transmission. Therefore, jittering needs to be handled properly to ensure system performance. 

Recently, a handful of studies in the existing literature have explored the uncertainty associated with UAV trajectories.
For example, the authors in \cite{UAVjitterintro} considered the jitter during UAV motion and the uncertainty of task size, 
and proposed a distributionally robust offloading and trajectory optimization algorithm to minimize the expected system delay.
In \cite{robustJitter}, the authors considered constraints for secrecy performance of the worst scenario and achieved energy-saving communication. However, the probability of the worst-case scenario occurring is usually very low.
The authors of \cite{robustJitter2} used the Gaussian distribution to model the uncertainty caused by UAV jittering and 
proposed a robust method for handling uncertainty in a probabilistic manner to achieve energy minimization.
In \cite{robustJitter3}, the authors innovatively designed an unsupervised learning approach to achieve robust trajectory design and rational resource allocation, with the goal of maximizing the minimum average spectral efficiency among mobile nodes. 
The above works on the UAV jittering mainly focused on robust trajectory optimization in the horizontal direction, 
but the influence of altitude changes is not mentioned. Additionally, data compression is not taken into account. 

In this paper, we investigate a three-dimensional robust trajectory and scheduling optimization scheme based on data compression. 
This scheme not only considers the real-world factor of UAV jittering, but also further adopts data compression technology to 
reduce the energy consumption of terminal users on the basis of trajectory planning and resource allocation.
In addition, the traditional UAV-assisted MEC model is difficult to meet the large-scale computing needs of terminal users, 
so we explore the cooperation model between UAV and ground base station (BS).
Specifically, MEC server is integrated into the BS to assist UAV in handling offloading 
tasks, which can alleviate the problem of insufficient computing resources for a single UAV.
To address the challenges imposed by the problem's non-convexity and the dynamic characteristics of the scene,
we employ the randomized ensembled double Q-learning (REDQ) algorithm in deep reinforcement learning (DRL) as a solution. 
This algorithm uses an update-to-data (UTD) ratio greater than 1, which enhances the model's data utilization efficiency.

The primary contributions of the paper are summarized as follows:
\begin{enumerate}
  \item We consider the trajectory uncertainty caused by UAV jittering in 3D space and, 
  based on this, propose an energy optimization scheme that integrates data compression technology. 
  This scheme can further reduce the energy consumption of terminal users by building on the joint optimization of trajectory planning and resource allocation.
  \item 
  The formulated problem is a non-convex optimization problem characterized by tightly coupled variables, 
  making it challenging to solve by using conventional optimization methods. 
  We reformulate the problem as a Markov Decision Process (MDP), 
  specifying the state space, action space, and reward function based on the given requirements. 
  We then introduce the REDQ algorithm to effectively solve the problem, and
  the proposed REDQ algorithm can adaptively optimize decisions in complex dynamic environments.
  \item We conduct a complexity analysis of the proposed algorithm and validate its performance through extensive simulation experiments. 
  The results show that the proposed algorithm significantly reduces the energy consumption of terminal users 
  while ensuring the robustness of UAV trajectories, outperforming existing optimization methods.
\end{enumerate}

The remainder of this paper is organized as follows. Section II introduces the system model and defines the problem. 
The REDQ algorithm is proposed in in Section III. 
Section IV evaluates the performance of the proposed scheme through simulation experiments
Finally, we make a summary in Section V. 

\section{SYSTEM MODEL AND PROBLEM FORMULATION}
\begin{figure}[t]
  \centering
  \includegraphics[width=\columnwidth]{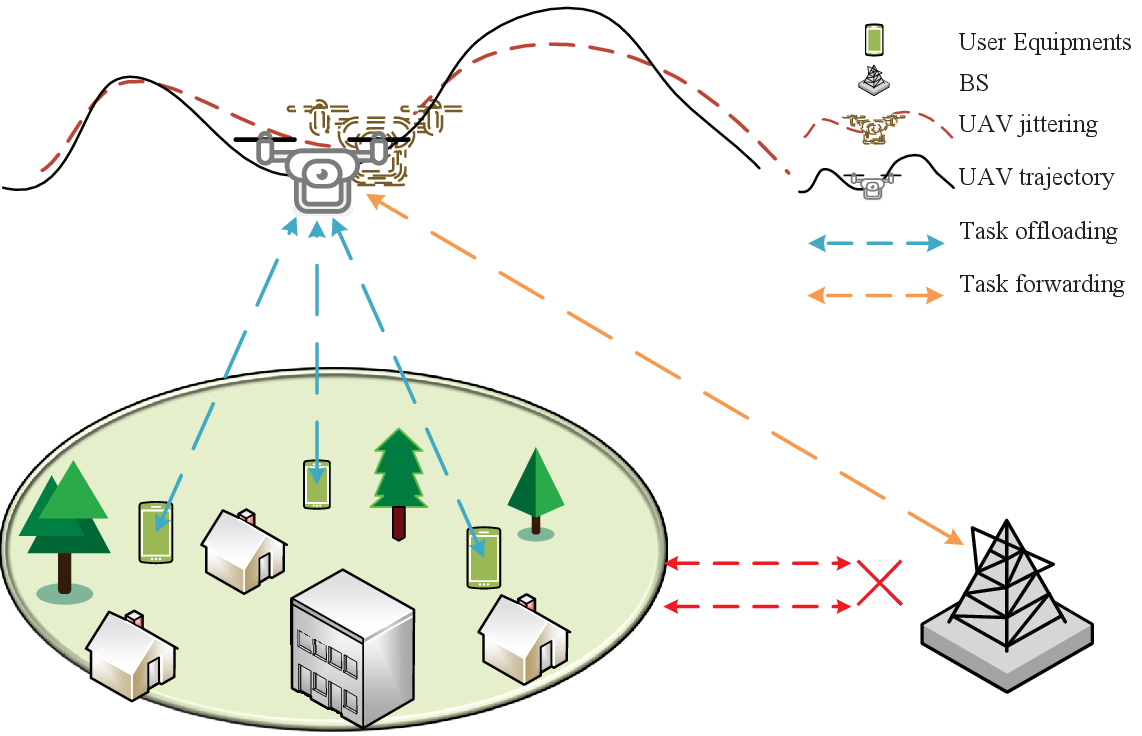}
  \caption{UAV-assisted MEC system with UAV jittering.}
  \label{fig:model}
\end{figure}

As shown in Fig. \ref{fig:model}, we consider a UAV-enabled MEC network, which comprises $K$ users, a UAV and a ground BS.
$K$ users are scattered within a area, and the set of users is denoted as $\mathcal{K}\triangleq\{1,2,\ldots,K\}$. The user $k\in \mathcal{K}$ coordinate is represented as ${\boldsymbol{w}_{k}}=[{{x}_{k}},{{y}_{k}},0]$.
The BS is equipped with abundant resources, and its location is denoted as ${\boldsymbol{w}_{\mathrm{BS}}}=[{{x}_{\mathrm{BS}}},{{y}_{\mathrm{BS}}},0]$.
The UAV is equipped with relay device and edge server, which can assist users in computing and serve as a medium for data transmission.
We assume that ground users will not directly offload tasks to the BS, due to low signal strength or poor channel conditions. When receiving the offloading tasks, UAV will further offload partial tasks to the BS for execution, due to limited computational capabilities and battery resources.

Assume the UAV's flight period is $T$, and it is divided into $N$ time slots equally.
The collection of time slots is represented as $\mathcal{N}\triangleq \{1,2,\ldots,N\}$, with each time slot having a duration of $\delta =T/N$.
At time instant $n$, the UAV is designed to take off from a specified starting point and 
travel to a specific location based on the user's task information to assist the terminal users in completing computing tasks. 
At these specific locations, the UAV will hover to facilitate data transmission and communication, 
ensuring stable connections with multiple terminal users. 
After completing the task, the UAV will fly to the next location in the next time slot to continue providing services to terminal users.
The UAV coordinate is defined as $\boldsymbol q[n]=[\boldsymbol L[n],H[n]]$, 
with $\boldsymbol L[n]=[x[n],\ y[n]]$ indicating the horizontal position and $H[n]$ indicating the UAV's altitude.
The UAV's displacement is restricted by its maximum speed and must adhere to the following constraints
\begin{align}
  & {{x}_{\min }}\le x[n]\le {{x}_{\max }},\ \forall n\in \mathcal{N},\label{regionx} \\ 
  & {{y}_{\min }}\le y[n]\le {{y}_{\max }},\ \forall n\in \mathcal{N},\label{regiony} \\
  & {{H}_{\min }}\le H[n]\le {{H}_{\max }},\ \forall n\in \mathcal{N},\label{regionh} \\
  & ||\boldsymbol q[n+1]-\boldsymbol q[n]||\le {{v}_{\max }}\delta ,
\end{align}
where ${{v}_{\max }}$ is the maximum flight speed. Constraints \eqref{regionx}, \eqref{regiony} and \eqref{regionh} denote the flight area restrictions for the UAV.

\subsection{Communication Model}
To avoid interference, we adopt orthogonal frequency division multiple access technology, and the total wireless bandwidth is divided equally.
The channels between the UAV and users, as well as between the UAV and the BS, are modeled as 
\begin{align}
  & {{h}_{k,\mathrm{u}}}[n]=\sqrt{{{\beta }_{0}}d_{k,\mathrm{u}}^{-\alpha }[n]}{{g}_{k,\mathrm{u}}},\forall n\in \mathcal{N},\\
  & {{h}_{\mathrm{u},\mathrm{BS}}}[n]=\sqrt{{{\beta }_{0}}d_{\mathrm{u},\mathrm{BS}}^{-\alpha }[n]}{{g}_{\mathrm{u},\mathrm{BS}}},\forall n\in \mathcal{N},
\end{align}
where ${{d}_{k,\mathrm{u}}}[n]=||\boldsymbol q[n]-{{\boldsymbol w}_{k}}||{}_{\text{2}}$ is the distance from the user $k$ to the UAV,
the distance between the UAV and the BS is denoted as ${{d}_{\mathrm{u},\mathrm{BS}}}[n]=||\boldsymbol q[n]-{{\boldsymbol w}_{\mathrm{BS}}}||{}_{\text{2}}$,
${{\beta }_{0}}$ is the strength gain value of the signal when the propagation distance is 1 meter,
$\alpha \ge 2$ represents the degree of loss during signal propagation,
and the small-scale fading ${{g}_{k,\mathrm{u}}}$ and ${{g}_{\mathrm{u},\mathrm{BS}}}$ are modelled as Rician fading, given by
\begin{align}
  & {{g}_{k,\mathrm{u}}}=\sqrt{\frac{{{K}_{k,\mathrm{u}}}}{{{K}_{k,\mathrm{u}}}+1}}g+\sqrt{\frac{1}{{{K}_{k,\mathrm{u}}}+1}}\tilde{g},\\
  & {{g}_{\mathrm{u},\mathrm{BS}}}=\sqrt{\frac{{{K}_{\mathrm{u},\mathrm{BS}}}}{{{K}_{\mathrm{u},\mathrm{BS}}}+1}}g+\sqrt{\frac{1}{{{K}_{\mathrm{u},\mathrm{BS}}}+1}}\tilde{g},
\end{align}
where $g$ represents the LOS component, whereas $\tilde{g}$ refers to the NLOS component that follows a Rayleigh distribution,
and ${{K}_{k,\mathrm{u}}}$ and ${{K}_{\mathrm{u},\mathrm{BS}}}$ are the Rician factors.
${{p}_{k}}$ and ${{p}_{\mathrm{u},k}}$ respectively denote the transmission power of the user and the UAV.
Then, the maximum achievable transmission rates from user $k$ to the UAV and the UAV to the BS are respectively given by
\begin{align}
  & {{r}_{k,\mathrm{u}}}[n]=B{{\log }_{2}}(1+\frac{{{p}_{k}}|{{h}_{k,\mathrm{u}}}[n]{{|}^{2}}}{B{{N}_{0}}}),\\
  & {{r}_{\mathrm{u},\mathrm{BS}}}[n]=B{{\log }_{2}}(1+\frac{{{p}_{\mathrm{u},k}}|{{h}_{\mathrm{u},\mathrm{BS}}}[n]{{|}^{2}}}{B{{N}_{0}}}),
\end{align}
where $B$ is the bandwidth resource allocated to users, and ${{N}_{0}}$ is the noise power distribution within a unit frequency range.
To avoid deriving complex cumulative distribution functions for random variables $|{{h}_{k,\mathrm{u}}}[n]{{|}^{2}}$ and $|{{h}_{\mathrm{u},\mathrm{BS}}}[n]{{|}^{2}}$, we approximate $|{{h}_{k,\mathrm{u}}}[n]{{|}^{2}}$ and $|{{h}_{\mathrm{u},\mathrm{BS}}}[n]{{|}^{2}}$ with logistic functions\cite{UAVRelayChannel}\cite{UAVRelayChannel2}.
Thus, ${{r}_{k,\mathrm{u}}}[n]$ and ${{r}_{\mathrm{u},\mathrm{BS}}}[n]$ can be rewritten as
\begin{align}
  & {{r}_{k,\mathrm{u}}}[n]={{B}_{k}}[n]{{\log }_{2}}(1+\frac{{{\beta }_{0}}{{p}_{k}}{{v}_{k,\mathrm{u}}}}{{{B}_{k}}[n]{{N}_{0}}{{(||q[n]-{{w}_{k}}|{{|}^{2}})}^{\alpha /2}}}),\\
  & {{r}_{\mathrm{u},\mathrm{BS}}}[n]={{B}_{k}}[n]{{\log }_{2}}(1+\frac{{{\beta }_{0}}{{p}_{\mathrm{u},k}}{{v}_{\mathrm{u},\mathrm{BS}}}}{{{B}_{k}}[n]{{N}_{0}}{{(||q[n]-{{w}_{\mathrm{BS}}}|{{|}^{2}})}^{\alpha /2}}}),
\end{align}
where ${{v}_{k,\mathrm{u}}}$ and ${{v}_{\mathrm{u},\mathrm{BS}}}$ denote the approximate fading power functions, and are defined as
\begin{align}
  & {{v}_{k,\mathrm{u}}}={{C}_{1}}+\frac{{{C}_{2}}}{1+{{e}^{-({{B}_{1}}+{{B}_{2}}{{u}_{k,\mathrm{u}}})}}},\\
  & {{v}_{\mathrm{u},\mathrm{BS}}}={{C}_{1}}+\frac{{{C}_{2}}}{1+{{e}^{-({{B}_{1}}+{{B}_{2}}{{u}_{\mathrm{u},\mathrm{BS}}})}}},
\end{align}
where ${{B}_{1}}<0$ denotes the positive logistic midpoint, ${{B}_{2}}>0$ is the logistic growth rate, 
${{C}_{1}}$ and ${{C}_{2}}$ are both greater than 0, and their sum equals 1,
and ${{u}_{k,\mathrm{u}}}=H[n]\text{/}{{d}_{k,\mathrm{u}}}[n]$ as well as ${{u}_{\mathrm{u},\mathrm{BS}}}=H[n]\text{/}{{d}_{\mathrm{u},\mathrm{BS}}}[n]$ represent the sine values of the elevation angle from the user to the UAV and from the UAV to the BS, respectively.
\subsection{Computation Model}
At the beginning of the $n$-th time slot, user $k$ generates a computing task ${{I}_{k}}[n]=\{{{D}_{k}}[n],{{C}_{k}}[n],t_{k}^{\max }\}$, where ${{D}_{k}}[n]$ denotes the amount of task data, ${{C}_{k}}[n]$ denotes the computing density, and $t_{k}^{\max }$ is the maximum tolerate delay of the task.
We assume that users do not perform task calculations and only perform data compression. The task is first compressed at the terminal user and then offloaded to the UAV, which processes partial task and further offloads the remaining portion to the BS for execution.
\subsubsection{Local compression}
In this paper, we adopt lossless compression technology and assume that all users use the same compression algorithm.
Let ${{\gamma }_{k}}[n]\in [{{\gamma }_{\min }},1]$ denote the task compression ratio, where ${{\gamma }_{\min }}$ is the minimum compression ratio.
Note that the mature model for assessing the computational complexity of data compression has not yet been established in the current literature.
A tractable model was proposed in reference \cite{compressionEqua}. 
The formula for the computational density of data compression can be defined as 
\begin{align}
  {{J}_{k}}[n]={{e}^{\epsilon /{{\gamma }_{k}}[n]}}-{{e}^{\epsilon }},
\end{align}
where $\epsilon$ is a constant associated with a specific compression algorithm.
When ${{\gamma }_{k}}[n]=1$, ${{J}_{k}}[n]=0$, which means that the compression algorithm is not executed.
Thus, the costs of local compression in terms of latency and energy are represented as
\begin{align}
  & {{t}_{k,\mathrm{lr}}}[n]=\frac{{{D}_{k}}[n]{{J}_{k}}[n]}{{{F}_{k}}},\\
  & {{e}_{k,\mathrm{lr}}}[n]={{D}_{k}}[n]{{J}_{k}}[n]F_{k}^{2}{{\tau }_{k}},
\end{align}
where ${{F}_{k}}$ represents the user's CPU frequency, and ${{\tau }_{k}}$ is the effective capacitance coefficient of the CPU, depending on the chip structure.
We assume that the terminal user has ample energy for data compression and offloading. In addition, the cost of decompression is much smaller than compression, so it is negligible\cite{UAVCompress}.
\subsubsection{Task offloading}
Task offloading comprises two stages: the first stage involves the UAV receiving tasks offloaded by the user, 
while the second stage involves the UAV further forwarding partial tasks to the BS.
To ensure the data integrity, we assume that the task offloading procedure will only be executed after the compression operation is completed.
The time and energy related to user offloading tasks are represented as follows
\begin{align}
  & {{t}_{k,\mathrm{off}}}[n]=\frac{{{\gamma }_{k}}[n]{{D}_{k}}[n]}{{{r}_{k,\mathrm{u}}}[n]},\\
  & {{e}_{k,\mathrm{off}}}[n]={{t}_{k,\mathrm{off}}}{{p}_{k}}.
\end{align}
The relay device of the UAV operates in half-duplex mode, and we assume that the UAV forwards the task only after completely receiving it, while the time for storing the task is ignored.
The costs of time delay and energy in UAV offloading can be defined as
\begin{align}
  & {{t}_{\mathrm{u},kr}}[n]=\frac{{{\alpha }_{k}}[n]{{\gamma }_{k}}[n]{{D}_{k}}[n]}{{{r}_{\mathrm{u},\mathrm{BS}}}[n]},\\
  & {{e}_{\mathrm{u},kr}}[n]={{t}_{\mathrm{u},kr}}{{p}_{\mathrm{u},k}},
\end{align}
where ${{\alpha }_{k}}[n]\in [0,1]$ is the task-partition ratio.
\subsubsection{Task computing for UAV}
${{F}_{\mathrm{u},k}}$ is a variable that needs to be optimized, representing the computation frequency allocated by the UAV to the user.
${{\tau }_{\mathrm{u}}}$ is the UAV's computing efficiency.
Therefore, the delay and energy costs for the UAV to process tasks offloaded by users are defined as follows
\begin{align}
  & t_{\mathrm{u},k}^{\mathrm{c}}[n]=\frac{(1-{{\alpha }_{k}}[n]){{C}_{k}}[n]{{D}_{k}}[n]}{{{F}_{\mathrm{u},k}}},\\
  & e_{\mathrm{u},k}^{\mathrm{c}}[n]=(1-{{\alpha }_{k}}[n]){{C}_{k}}[n]{{D}_{k}}[n]{{\tau }_{\mathrm{u}}}F_{\mathrm{u},k}^{2}.
\end{align}
\subsubsection{Task computing for BS}
Due to the sufficient energy supply and abundant computing resources of the BS, we only consider the computation latency at the BS. The latency consumed by edge servers in computing offloading tasks can be expressed as
\begin{align}
  t_{\mathrm{BS},k}^{\mathrm{c}}[n]=\frac{{{\alpha }_{k}}[n]{{C}_{k}}[n]{{D}_{k}}[n]}{{{F}_{\mathrm{BS}}}},
\end{align}
where ${{F}_{\mathrm{BS}}}$ represents the computing resources of the BS.

The UAV energy consumption model considers three components: flight ${{e}_{\mathrm{fly}}}[n]$, computation $e_{\mathrm{u},k}^{c}[n]$, and forwarding ${{e}_{\mathrm{u},kr}}[n]$.
In time slot $n$, the total energy consumption of the UAV is expressed as
\begin{align}
  \begin{split}
    {{E}_{\mathrm{u},\mathrm{sum}}}[n]=\sum\limits_{k=1}^{K}{{{e}_{\mathrm{u},kr}}[n]+e_{\mathrm{u},k}^{c}[n]+{{e}_{\mathrm{fly}}}[n]},
  \end{split}
\end{align}
where the flight energy consumption is given by ${{e}_{\mathrm{fly}}}[n]=\delta {{p}_{\mathrm{fly}}}$.
The propulsion power of the rotary-wing UAV during $n$-th time slot, as outlined in \cite{UAVEnergy}, can be denoted as
\begin{equation}
  \begin{split}
    {{p}_{\mathrm{fly}}}=& {{p}_{0}}\left( 1+\frac{3{{\left( {{v}^{\mathrm{h}}}[n] \right)}^{2}}}{U_{\mathrm{tip}}^{2}} \right)+\frac{1}{2}{{d}_{0}}\rho sG{{\left( {{v}^{\mathrm{h}}}[n] \right)}^{3}}+\\
    & {{p}_{1}}{{\left( \sqrt{1+\frac{{{\left( {{v}^{\mathrm{h}}}[n] \right)}^{4}}}{4v_{0}^{4}}}-\frac{{{\left( {{v}^{\mathrm{h}}}[n] \right)}^{2}}}{2v_{0}^{2}} \right)}^{\tfrac{1}{2}}}+{{p}_{2}}{{v}^{\mathrm{v}}}[n],
  \end{split}
\end{equation}
where ${{v}^{\mathrm{h}}}[n]={{\left\| \boldsymbol L[n+1]- \boldsymbol L[n] \right\|}_{2}}\text{/}\delta$ denotes the UAV's horizontal flight 
speed, ${{v}^{\mathrm{v}}}[n]={{\left\| H[n+1]-H[n] \right\|}_{2}}\text{/}\delta$ represents the UAV's vertical speed,
${{p}_{0}}$ is the blade profile power, ${{p}_{1}}$ is the induction power of UAV in hovering state, ${{p}_{2}}$ denotes the decreasing or increasing power,
the rotor blade's tip speed is defined as ${{U}_{\mathrm{tip}}}$, ${{v}_{0}}$ is the average rotor induced speed when the UAV is hovering,  
the fuselage drag ratio is expressed as ${{d}_{0}}$, $\rho$ signifies the air density, $s$ denotes the rotor disc area, and $G$ is the rotor solidity.

The computation module and communication module on the UAV are usually separate, so computation can be carried out simultaneously with task transmission.
During $n$-th time slot, the total latency consumed by user $k$ to complete a computing task can be expressed as
\begin{align}
  \begin{split}
    {{t}_{k}}[n]=& {{t}_{k,\mathrm{lr}}}[n]+{{t}_{k,\mathrm{off}}}[n]+\\
    & \max \{t_{\mathrm{u},k}^{\mathrm{c}}[n],{{t}_{\mathrm{u},kr}}[n]+t_{\mathrm{BS},k}^{\mathrm{c}}[n]\}.
  \end{split}
\end{align}
The energy cost for user $k$ consists of two parts: compression and offloading, specifically formulated as follows
\begin{align}
{{e}_{k}}[n]={{e}_{k,\mathrm{lr}}}[n]+{{e}_{k,\mathrm{off}}}[n].
\end{align}
\subsection{Uncertainty Model}
In actual scenarios, due to inaccurate positioning information, imperfect flight control, air turbulence and other factors, the UAV may deviate from the scheduled trajectory, which will affect the key performance of edge computing network, such as energy consumption, time delay and computing efficiency.
Therefore, in order to guarantee the performance of the system in the actual operating environment, unpredictable UAV trajectory caused by uncertainties need to be specially addressed to design a robust high-efficient UAV-assisted MEC system.
The uncertainty trajectory can be modeled as
\begin{align}
  \boldsymbol q[n]=\hat{\boldsymbol q}[n]+\Delta \boldsymbol q[n],\ \forall n\in \mathcal{N},
\end{align}
where $\hat{\boldsymbol q}[n]$ is the planned trajectory, $\Delta \boldsymbol q[n]$ is the position error caused by UAV jittering.
Inspired by \cite{robustJitter2}, 
the uncertainty can be modeled as a Gaussian random variable, defined as follows
\begin{align}
  \Delta \boldsymbol q[n]\sim \mathcal{N}(0,\varepsilon _{0}^{2} \boldsymbol I),\,\ \forall n\in \mathcal{N},
\end{align}
where $\boldsymbol I$ is the third-order identity matrix, corresponding to the three-dimensional coordinate axis.
Due to the UAV trajectory being modeled as a random variable, the constraints on the UAV trajectory mentioned above need to be modified
\begin{align}
  {{\mathcal{P}}_{\Delta \boldsymbol q[n]}}\{||\boldsymbol q[n+1]-\boldsymbol q[n]||\le {{v}_{\max }}\delta \}\ge 1-\rho _{n}^{(\mathrm{trj})},\forall n\in \mathcal{N},
\end{align}
where $\rho _{n}^{(\mathrm{trj})}$ defines the probability that the UAV's flight speed exceeds the maximum flight speed, 
which is derived based on uncertainty.
\subsection{Problem Formulation}
For UAV-enabled MEC network, we propose an optimization problem aimed at minimizing the total cost of energy for all terminal users through joint 
optimization of UAV trajectory $\boldsymbol q=\{\boldsymbol q[n],\forall n\in \mathcal{N}\}$, 
data compression ratios $\boldsymbol \gamma =\{{{\gamma }_{k}}[n],\forall k\in \mathcal{K},\forall n\in \mathcal{N}\}$, 
computing resource allocation ${{\boldsymbol F}_{\mathrm{u}}}=\{{{F}_{\mathrm{u},k}}[n],\forall k\in \mathcal{K},\forall n\in \mathcal{N}\}$, 
and task offloading ratios $\boldsymbol \alpha =\{{{\alpha }_{k}}[n],\forall k\in \mathcal{K},\forall n\in \mathcal{N}\}$.
The specific optimization problem is formulated as follows
\begin{subequations}\label{P:0}
  \begin{align}
    & \underset{\boldsymbol q,\boldsymbol \alpha ,\boldsymbol \gamma ,{{\boldsymbol F}_{\mathrm{u}}}}{\mathop{\min }}\,\quad \sum\limits_{n=1}^{N}{\sum\limits_{k=1}^{K}{{{e}_{k}}[n]}} \label{P0:Obj} \\
    \text{s.t.}~
    & {{x}_{\min }}\le x[n]\le {{x}_{\max }},\ \forall n\in \mathcal{N} \label{P0:x} \\
    & {{y}_{\min }}\le y[n]\le {{y}_{\max }},\ \forall n\in \mathcal{N} \label{P0:y} \\
    & {{H}_{\min }}\le H[n]\le {{H}_{\max }},\forall n\in \mathcal{N} \label{P0:h}\\
    & \mathcal{P}\{||\boldsymbol q[n+1]-\boldsymbol q[n]||\le {{v}_{\max }}\delta \}\ge 1-\rho _{n}^{(\mathrm{trj})},\forall n\in \mathcal{N} \label{P0:dis}\\
    & \sum\limits_{n=1}^{N}{\sum\limits_{k=1}^{K}{{{e}_{\mathrm{u},kr}}[n]}}+e_{\mathrm{u},k}^{\mathrm{c}}[n]+{{e}_{\mathrm{fly}}}[n]\le {{E}_{\mathrm{u},\max }} \label{P0:Euav}\\
    & {{\alpha }_{k}}[n]\in [0,1],\forall n\in \mathcal{N} \label{P0:off}\\
    & {{\gamma }_{k}}[n]\in [{{\gamma }_{\min }},1],\forall n\in \mathcal{N} \label{P0:com}\\
    & \sum\limits_{k=1}^{K}{{{F}_{\mathrm{u},k}}[n]\le {{F}_{\mathrm{u},\max }},}\forall n\in \mathcal{N} \label{P0:Fsum}\\
    & 0\le {{F}_{\mathrm{u},k}}[n]\le {{F}_{\mathrm{u},\max }},\forall k\in \mathcal{K},\forall n\in \mathcal{N} \label{P0:Fk}\\
    & {{t}_{k}}[n]\le t_{k}^{\max },\forall k\in \mathcal{K},\forall n\in \mathcal{N} \label{P0:tmax}
  \end{align}
\end{subequations}
where ${{E}_{\mathrm{u},\max }}$ represents the UAV's total energy, and ${{F}_{\mathrm{u},\max }}$ denotes total computing resources. 
Constraint \eqref{P0:x} and constraint \eqref{P0:y} represent the flight range limitations of the UAV.
Constraint \eqref{P0:h} reflects the altitude limit for the UAV flight.
Constraint \eqref{P0:dis} is the displacement limit within adjacent time slots of the UAV.
Constraint \eqref{P0:Euav} specifies that the energy cost incurred by the UAV must not surpass the maximum limit.
Constraint \eqref{P0:off} and constraint \eqref{P0:com} represent the offloading ratio and data compression ratio, respectively.
Constraint \eqref{P0:Fsum} and constraint \eqref{P0:Fk} reflect the UAV's computing resource constraints.
Constraint \eqref{P0:tmax} represents the latency requirement for task calculation.

\section{DRL-BASED ALGORITHM}
The optimization problem mentioned above is a non-convex problem with highly coupled variables and involves a long-term energy minimization objective, which is difficult to solve using traditional optimization algorithms quickly and effectively. 
As a promising solution, DRL is particularly well-suited for such long-term optimization problems. 
Guided by an appropriate reward function, the DRL agent continuously interacts with the environment, 
learning the dynamic characteristics of the system to provide high-quality decision-making solutions \cite{DRL1}. 
Furthermore, the real-time decision-making capability of DRL allows the agent to flexibly adjust its strategy based on changes in the environment, responding promptly to the system's needs. 
This dynamic adjustment not only enhances the adaptability of the decisions but also ensures that the agent can 
continuously optimize system performance in complex and changing environments.

\subsection{Analysis of Model-Free MDP}
The optimization problem is commonly converted into an MDP to enable its solution with DRL algorithms.
In model-free DRL algorithms, the MDP is characterized by the tuple $(\mathcal{S},\mathcal{A},\mathcal{R},\gamma )$, 
where $\mathcal{S}$ stands for the state space, $\mathcal{A}$ indicates the action space, the reward function is represented as $\mathcal{R}$, 
and $\gamma$ is the discount factor. 
$\gamma \in [0,1]$ is defined as a constant used to measure the relative importance of future rewards.
The definitions of the other three elements are as follows:
\subsubsection{State Space}State is the basis for the agent to make decisions. We define the state ${{\boldsymbol s}_{n}}\in \mathcal{S}$ during the $n$-th time slot as
\begin{align}
  {{\boldsymbol s}_{n}}=\text{ }\!\!\{\!\!\text{ } \boldsymbol q[n],{{E}_{\mathrm{battery}}},{{D}_{k}}[n],{{C}_{k}}[n],\forall k\in \mathcal{K}\text{ }\!\!\}\!\!\text{ },
\end{align}
where ${{E}_{\mathrm{battery}}}$ denotes the UAV's remaining power. The dimensionality of the state space is determined by the user information and the UAV's state information. 
Specifically, the user's task information includes the task size and computation density, 
which directly impacts the energy and latency-related costs.
The UAV's state includes its position and battery level, which can indirectly affect task offloading by influencing the communication link.
Therefore, the dimensionality of the state space is defined as $2k+4$.
Moreover, there are significant differences in the magnitude of different state values in the state, which can affect the training of the agent. Thus, we preprocess the state values and scale them to the range of 0 to 1.

\subsubsection{Action Space}Action is the result of the agent decision-making, which reflects optimization variables and can be defined as
\begin{align}
  {{\boldsymbol a}_{n}}=\{\bar{\boldsymbol q}[n],{{F}_{\mathrm{u},k}}[n],{{\alpha }_{k}}[n],{{\gamma }_{k}}[n],\forall k\in \mathcal{K}\text{ }\!\!\}\!\!\text{ },
\end{align}
where $\bar{\boldsymbol q}[n]$ indicates the UAV's movement distance.
The action space dimensionality is determined by both the users and the UAV. 
Each user's action includes the offloading ratio, compression ratio, 
and resource allocation ratio, while the UAV's action involves three-dimensional trajectory decisions. 
Therefore, the total action space dimensionality is $3k + 3$.
Contrary to state value preprocessing, we set the output of the agent within the range of 0 to 1, and then restore it to the actual value through a mapping function.

\subsubsection{Reward Formulation}The reward function is used to guide the agent to take the correct action.
The reward function in this paper consists of two parts: optimization objective and punishment, which can be defined as follows
\begin{align}
  r[n]=-{{E}_{\mathrm{sum}}}[n]{{P}_{\mathrm{T}}}[n]{{P}_{\mathrm{E}}}[n]{{P}_{\Delta \boldsymbol q}}[n],
\end{align}
where ${{E}_{\mathrm{sum}}}[n]=\sum{_{k\in \mathcal{K}}{{e}_{k}}[n]}$ is all terminal users' energy consumption  
at time instant $n$, ${{P}_{\Delta \boldsymbol q}}[n]$, ${{P}_{\mathrm{E}}}[n]$ and ${{P}_{\mathrm{T}}}[n]$ 
denote the penalties for not meeting the constraint \eqref{P0:dis}, the constraint \eqref{P0:Euav} and the constraint \eqref{P0:tmax}, respectively. 
We adopt the penalty function $P(x,a,b)=2-\exp (-{{\left[ (x-a)/b \right]}^{+}})$ to calculate the penalty value. 
${{P}_{\Delta \boldsymbol q}}[n]$, ${{P}_{\mathrm{E}}}[n]$ as well as ${{P}_{\mathrm{T}}}[n]$ can be described as follows
\begin{align}
  & {{P}_{\Delta \boldsymbol q}}[n]=P(\text{1}-{{\mathcal{P}}_{\Delta \boldsymbol q[n]}},\ \rho _{n}^{(\mathrm{trj})},\ \rho _{n}^{(\mathrm{trj})}), \\
  & {{P}_{\mathrm{E}}}[n]=P({{E}_{\mathrm{u},\mathrm{cum}}}[n],{{E}_{\mathrm{u},\max }},{{E}_{\mathrm{u},\max }}), \\
  & {{P}_{\mathrm{T}}}[n]=\frac{1}{K}\sum\limits_{k=1}^{K}{P({{t}_{k}}[n],{{t}_{k,\max }},{{t}_{k,\max }})},
\end{align}
where ${{E}_{\mathrm{u},\mathrm{cum}}}[n]$ denotes the cumulative energy expenditure of the UAV over the first $n$ time slots and can be defined as
\begin{align}
  {{E}_{\mathrm{u},\mathrm{cum}}}[n]=\sum\limits_{i=1}^{n}{\sum\limits_{k=1}^{K}{{{e}_{\mathrm{u},kr}}[i]}}+e_{\mathrm{u},k}^{\mathrm{c}}[i]+{{e}_{\mathrm{fly}}}[i].
\end{align}
\subsection{REDQ-Based Optimization Algorithm}
Unlike other machine learning algorithms, the agent of DRL seeks the optimal strategy by interacting with the environment without requiring any prior information about the environment, 
which makes DRL highly adaptable in dealing with complex and uncertain environments.
The DRL algorithms can be divided into two categories based on the type of action space: discrete action space algorithms and continuous action space algorithms.
The typical algorithms for discrete action space include DQN, DDQN, and so on, while algorithms for continuous action space include PPO, DDPG, etc.
The optimization variables in the formulated problem, such as offloading ratio and computing resource allocation, are continuous, so the discrete action space algorithm is not suitable.
Compared to other continuous action space algorithms, the REDQ algorithm \cite{redq} has the following advantages: 
1) The algorithm adopts a stochastic policy to output the probability distribution of actions, which is more conducive to exploring action space compared to deterministic policy algorithms; 2) The algorithm belongs to the off-policy algorithms and has a higher sample efficiency rate compared to the on-policy algorithms.
\begin{figure}[t]
  \centering
  \includegraphics[width=\columnwidth]{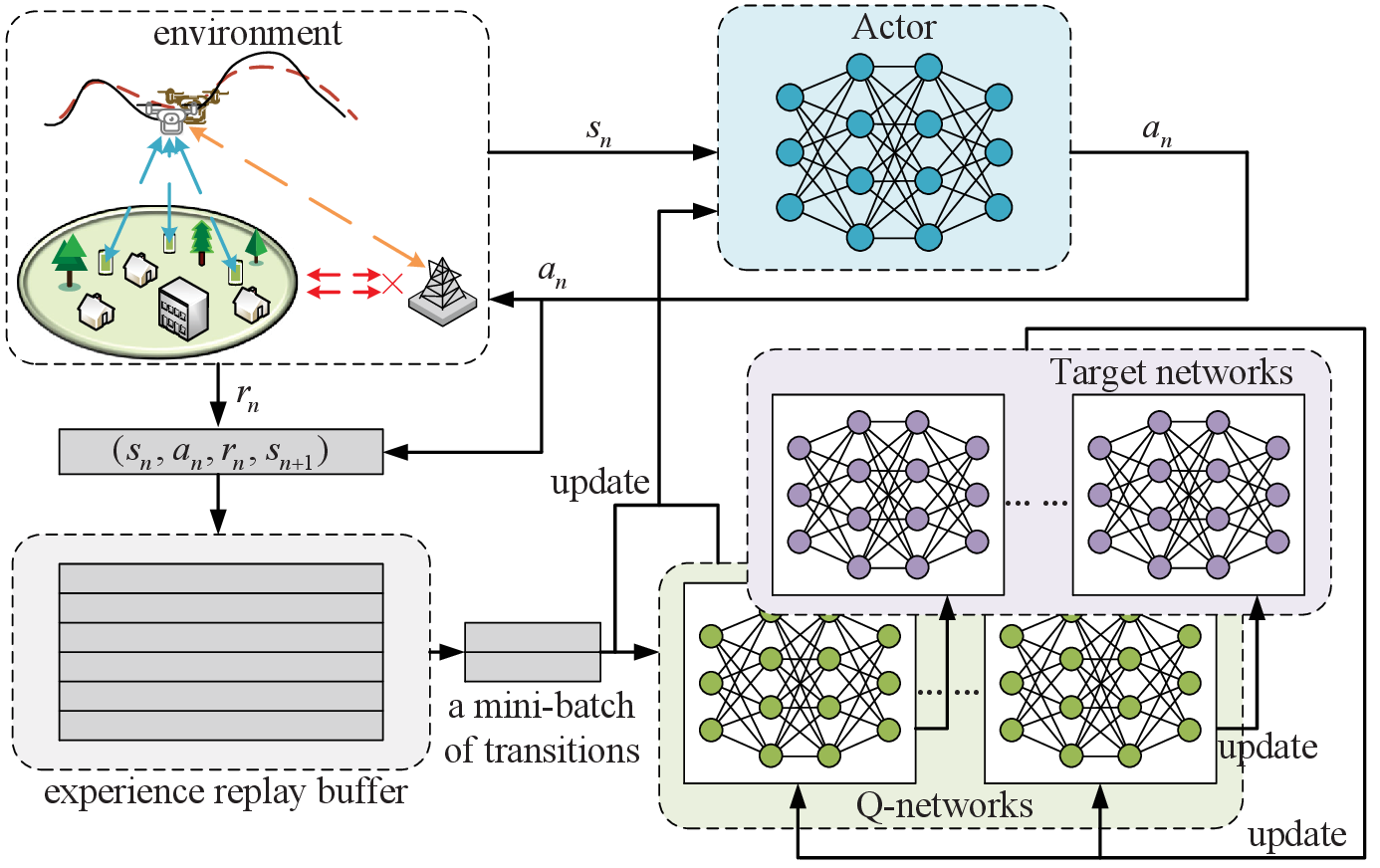}
  \caption{Architecture diagram of the REDQ algorithm.}
  \label{fig:REDQ}
\end{figure}

The REDQ algorithm builds upon the Soft Actor-Critic algorithm and incorporates the following two key components:
1) introduce $X$ Q-functions to reduce variance and stabilize the training process;
2) when calculating the Q target value, randomly select a subset $\mathcal{Y}$ of the $X$ Q target networks 
and take minimum value from the subset to alleviate the overestimation problem.
The above two critical components enable the REDQ algorithm to maintain stable and nearly uniform bias under high UTD ratio.
The REDQ algorithm is founded on the maximum entropy principle, aiming to optimize 
both cumulative reward and the policy's entropy.
${{\pi }^{*}}$ denotes the optimal strategy, defined as follows
\begin{align}
  {{\pi }^{*}}=\arg \underset{\pi }{\mathop{\max }}\,E\left[ \sum\limits_{n=0}^{T}{{{\gamma }^{n}}({{r}_{n}}({{\boldsymbol s}_{n}},{{\boldsymbol a}_{n}})}+\alpha H(\pi (\cdot |{{\boldsymbol s}_{n}}))) \right],
\end{align}
where $\alpha$ is a weight parameter which balances the importance between entropy and the reward, 
and the entropy $H(\pi (\cdot |{{\boldsymbol s}_{n}}))$ is defined as follows
\begin{align}
  H(\pi (\cdot |{{\boldsymbol s}_{n}}))={{E}_{{{\boldsymbol a}_{n}}}}[-\log \pi ({{\boldsymbol a}_{n}}|{{\boldsymbol s}_{n}})].
\end{align}
Slightly different from conventional DRL algorithms, the Q-function and V-function in this paper, incorporating the entropy term, are defined as follows
\begin{align}
  {{V}^{\pi }}(\boldsymbol s)=E\left[ \sum\limits_{n=0}^{T}{{{\gamma }^{n}}\left( {{r}_{n}}+\alpha H(\pi (\cdot |{{\boldsymbol s}_{n}})) \right)}\left| {{\boldsymbol s}_{0}}=\boldsymbol s \right. \right], 
\end{align}
\begin{align}
  \begin{split}
    {{Q}^{\pi }}(\boldsymbol s,\boldsymbol a)=& E\left[ \sum\limits_{n=0}^{T}{{{\gamma }^{n}}{{r}_{n}}}+ \right.\\
    & \left. \alpha \sum\limits_{n=1}^{T}{{{\gamma }^{n}}H(\pi (\cdot |{{\boldsymbol s}_{n}}))}\left| {{\boldsymbol s}_{0}}=\boldsymbol s,{{\boldsymbol a}_{0}}=\boldsymbol a \right. \right].
  \end{split}
\end{align}
The training framework of the REDQ algorithm is shown in Fig. \ref{fig:REDQ}, 
where the agent consists of a actor network, $X$ critic networks, and $X$ target networks.
The Actor network ${{\pi }_{\phi }}(\ \cdot \ |{{\boldsymbol s}_{n}})$ is used to approximate the policy function, where $\phi$ denotes the parameter. 
The critic network ${{Q}_{{{\theta }_{i}}}}({{\boldsymbol s}_{n}},{{\boldsymbol a}_{n}})$ is used to approximate action-value function, 
parameterized by ${{\theta }_{i}}$, and the parameter of the target $Q$-network ${{Q}_{{{{\bar{\theta }}}_{i}}}}({{\boldsymbol s}_{n}},{{\boldsymbol a}_{n}})$ is represented as ${{\bar{\theta }}_{i}}$, $i=1,2,3,...,X$.
The algorithm adopted in this paper is an off-policy algorithm, 
which allows for the separation of training and data collection processes. 
The agent first uses the Actor network to determine the action based on the current system state at each time step.
After the UAV and terminal users perform corresponding actions, the system environment will provide feedback results and update the state.
The agent stores the interaction records in the experience replay pool $\mathcal{D}$ and, when computing resources are available, 
randomly samples a portion $\mathcal{B}$ of these samples to train and update the parameters of three different neural networks.
Due to the large amount of computational resources required during the training process, 
we deploy the offline training process on the BS with abundant computing resources.
Specifically, each $Q$ function is updated using a shared target, which is defined as follows
\begin{align}
  y={{r}_{n}}+\gamma \left( \underset{i\in \mathcal{Y}}{\mathop{\min }}\,{{Q}_{{{{\bar{\theta }}}_{i}}}}({{\boldsymbol s}_{n+1}},{{\boldsymbol a}_{n+1}})-\alpha \log {{\pi }_{\phi }}({{\boldsymbol a}_{n+1}}|{{\boldsymbol s}_{n+1}}) \right), \label{targety}
\end{align}
where ${{\boldsymbol a}_{n+1}}\sim {{\pi }_{\phi }}(\cdot |{{\boldsymbol s}_{n+1}})$, 
and the elements in set $\mathcal{Y}$ are sampled from the $X$ target networks.
We use gradient descent to adjust the parameters of the networks,
with the gradients defined as follows
\begin{align}
  \begin{split}
    \nabla {{L}_{\pi }}(\phi )=& {{\nabla }_{\phi }}\frac{1}{|\mathcal{B}|}\\
    & \sum\limits_{{{\boldsymbol s}_{n}}\in \mathcal{B}}{\left( \frac{1}{X}\sum\limits_{i=1}^{X}{{{Q}_{{{\theta }_{i}}}}({{\boldsymbol s}_{n}},{{{\tilde{\boldsymbol a}}}_{n}})-\alpha \log {{\pi }_{\phi }}({{{\tilde{\boldsymbol a}}}_{n}}|{{\boldsymbol s}_{n}})} \right)}, \label{losspi}
  \end{split}
\end{align}
\begin{align}
  \nabla {{L}_{Q}}({{\theta }_{i}})={{\nabla }_{\theta }}\frac{1}{|\mathcal{B}|}{{\sum\limits_{\left( {{\boldsymbol s}_{n}},{{\boldsymbol a}_{n}},{{r}_{n}},{{\boldsymbol s}_{n+1}} \right)\in \mathcal{B}}{\left( {{Q}_{{{\theta }_{i}}}}({{\boldsymbol s}_{n}},{{\boldsymbol a}_{n}})-y \right)}}^{2}}, \label{lossQ}
\end{align}
where ${{\tilde{\boldsymbol a}}_{n}}\sim {{\pi }_{\phi }}(\cdot |{{\boldsymbol s}_{n}})$.
The target network adopts a soft-update, which updates the parameters of the target network by slowly tracking the Critic network parameters and is defined as follows
\begin{align}
  {{\bar{\theta }}_{i}}\leftarrow \tau {{\bar{\theta }}_{i}}+(1-\tau ){{\theta }_{i}}, \label{targetQ}
\end{align}
where $\tau$ is the learning rate. We adopt an adaptive gradient-based method to adjust the entropy weight.
When the agent explores the new space, the optimal action is not yet clear, and the weight should be increased to encourage exploration.
As the optimal action is gradually determined, it will be gradually decreased.
The weight can be updated through the following loss function
\begin{align}
  L(\alpha )={{E}_{{{\boldsymbol a}_{n}}\sim{{\pi }_{\phi }}}}[-\alpha \log {{\pi }_{\phi }}({{\boldsymbol a}_{n}}|{{\boldsymbol s}_{n}})-\alpha \bar{H}], \label{lossentropy}
\end{align}
where $\bar{H}$ is the predefined minimum entropy threshold.
Based on the above description, the detailed REDQ-based total energy consumption minimization algorithm is illustrated in Algorithm 1.

\subsection{Computational Complexity Analysis}
The computational complexity of the proposed algorithm is closely related to the structure of the neural networks and the number of neurons. 
Specifically, the complexity analysis arises from two stages: training and execution.
During the training phase, 
both the Actor and Critic networks typically require multiple iterations, 
with each iteration involving gradient calculations and parameter updates. 
The complexity is represented as $O({{N}_{\mathcal{B}}}(\sum\nolimits_{i=\text{0}}^{I-1}{{{l}_{i}}{{l}_{i+1}}}\text{+}\sum\nolimits_{j=0}^{J-1}{{{{\hat{l}}}_{j}}{{{\hat{l}}}_{j+1}}}))$,
where ${{N}_{\mathcal{B}}}$ denotes the size of the mini-batch, $I$ and $J$ define the number of fully connected layers in the Actor and Critic, respectively.
Moreover, the numbers of neurons in the $i$-th layer of the Actor network and the $j$-th layer of the Critic network are represented as ${{l}_{i}}$ and ${{\hat{l}}_{j}}$, respectively.
During the execution process, only the Actor network that has been trained is required for application.
Therefore, the computational complexity is equivalent to the cost of performing forward propagation through the Actor network, 
which is defined as $O(\sum\nolimits_{i=\text{0}}^{I-1}{{{l}_{i}}{{l}_{i+1}}})$.

\begin{algorithm}[t]
  \caption{Training framework of the REDQ}
  \label{redq}
  \begin{algorithmic}[1]
      \STATE{Initialize Actor network parameters $\phi$, Critic network parameters ${{\theta }_{i}}$, $i=1,2,3,...,X$, 
      experience replay buffer $\mathcal{D}$. Set target parameters ${{\bar{\theta }}_{i}}\leftarrow {{\theta }_{i}}$, 
      for ${{\theta }_{i}}$, $i=1,2,3,...,X$.}
      \FOR{step $s=1,2,...,Ms$}
          
          \STATE{Obtain state ${{\boldsymbol s}_{n}}$ from the environment.}
          \STATE{Sample action: ${{\boldsymbol a}_{n}}\sim {{\pi }_{\phi }}(\cdot |{{\boldsymbol s}_{n}})$ and execute actions ${{\boldsymbol a}_{n}}$.}
          \STATE{Obtain feedback ${{r}_{n}}$ and observe new state ${{\boldsymbol s}_{n+1}}$.}
          \STATE{Add data to buffer: $\mathcal{D}\leftarrow \mathcal{D}\cup \text{ }\!\!\{\!\!\text{ }{{\boldsymbol s}_{n}}\text{,}\ {{\boldsymbol a}_{n}}\text{,}\ {{r}_{n}}\text{,}\ {{\boldsymbol s}_{n+1}}\}$.}
        
          \FOR{update $u=1,2,...,Mu$}
              \STATE{Sample a min-batch $\mathcal{B}$ from experience replay buffer $\mathcal{D}$.}
              \STATE{Sample a set $\mathcal{Y}$ from $\{1,\ 2,\ ...,\ X\}$.}
              \STATE{Calculate the target value according to \eqref{targety}.}
              \FOR{${{\theta }_{i}}$, $i=1,2,3,...,X$}
                  \STATE{Update critic network parameters according to equation \eqref{lossQ}.}
                  \STATE{Update target network parameters with equation \eqref{targetQ}.}
              \ENDFOR   
          \ENDFOR
          \STATE{Update Actor network parameters with equation \eqref{losspi}.}
          \STATE{Update the entropy weight with equation \eqref{lossentropy}.}
      \ENDFOR 
  \end{algorithmic}
\end{algorithm}

\section{SIMULATION RESULTS AND DISCUSSION}
In this section, we assess the performance of the proposed algorithm using a series of comprehensive experiments.
We set the horizontal area of the simulation environment to a rectangular area of $500\ \mathrm{m}$ $\times$ $500\ \mathrm{m}$, 
within which users and the UAV are located.
In the vertical direction, users are scattered on the ground while the UAV's altitude is configured within the range of [$100$, $200$] $\mathrm{m}$.
The BS is located at [$500$, $500$, $0$] $\mathrm{m}$.
The user set size is configured to $15$, and the data size of task is uniformly distributed within [$1$, $2.5$] $\mathrm{Mbits}$.
The computing density is set to [$700$, $1000$] $\mathrm{cycles}/\mathrm{bit}$.
For simplicity, we set the maximum allowable delay to the duration of the time slot.
The onboard energy of the UAV is set to $20000$ $\mathrm{J}$, the starting position is set to [$0$, $0$, $150$] $\mathrm{m}$, 
and the maximum flight speed is set to $20$ ${\mathrm{m}}/{\mathrm{s}}$.
Other experimental parameters are set according to \cite{UAVRelayChannel} and \cite{UAVEnergy}.

The networks in this paper are constructed using fully connected layers with ReLU activation functions, featuring $2$ hidden layers of $128$ neurons each.
The set size of the value networks is set to $10$, and its subset size is set to $2$.
The discount coefficient is $0.9$, the experience replay buffer is set to $20000$, and the training sample set size is configured to $256$.
The learning rates are set to $0.0001$ for the policy network and $0.001$ for the value network, respectively.
We use the adam optimizer to update the model parameters.
\begin{table}[htp]
  \centering
  \caption{Simulation Parameters.} \label{Simulation}
  \renewcommand\arraystretch{1.5}
  \tabcolsep=0.3cm
  \begin{tabular}{c|c}
    \hline
    \textbf{Parameters}                    & \textbf{Values}   \\ \hline
    Number of time slots                   &   $50$            \\
    Length of time slot                    &   $1.5$ $\mathrm{s}$        \\
    CPU frequency of user                  &   $1.5$ $\mathrm{GHz}$      \\
    CPU frequency of UAV                   &   $30$ $\mathrm{GHz}$       \\
    CPU frequency allocated by BS to user  &   $15$ $\mathrm{GHz}$       \\
    Bandwidth resource                     &   $30$ $\mathrm{GHz}$       \\
    Minimum data compression ratio         &   $0.5$           \\
    Capacitance coefficient                &   ${{10}^{-29}}$  \\
    Probability of speed violation         &   $0.1$           \\
    Background noise power spectrum density&   $-174$ $\mathrm{dBm}/\mathrm{Hz}$ \\ \hline
  \end{tabular}
\end{table}

To assess the effectiveness of the proposed REDQ algorithm, we conducted a comparison with the following algorithms.
\begin{itemize}
  \item {\bf{PPO:}} PPO is an on-policy algorithm that adopts a stochastic policy. 
  Gaussian distribution is utilized to model the probability distribution of actions, where the Actor network outputs mean, and the standard deviation is a constant. 
  \item {\bf{A2C:}} The algorithm combines the advantages of policy gradients and value functions, which alleviates the high variance problem 
  and stabilizes the training process by introducing an advantage function.
  \item {\bf{Random move:}} The trajectory of the UAV are chosen randomly while other actions are determined using the proposed algorithm. 
  This approach is employed to assess the effect of trajectory planning on system performance.
  \item {\bf{Untreated scheme:}} The algorithm assumes the presence of UAV jittering but does not address it,
   aiming to highlight the impact of jitter on performance and validate the robustness of the proposed solution.
\end{itemize}

\begin{figure}[t]
  \centering
  \includegraphics[width=7.7cm]{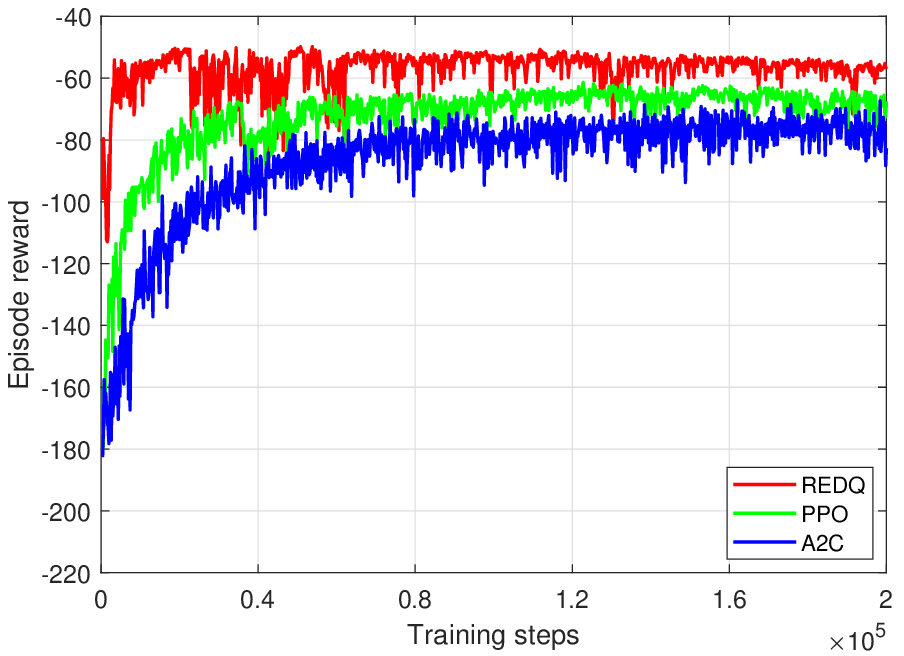}
  \caption{Convergence performances of different algorithms.}
  \label{fig:Convergence}
\end{figure}
Fig. \ref{fig:Convergence} shows the convergence behavior of different algorithms. It is evident that with an increase in training steps, 
the convergence values for all schemes steadily decline.
This indicates that through interaction with the environment, the agent can learn effective policy, 
thereby achieving the goal of reducing energy consumption.
When the training steps reach about $\text{4}\times \text{1}{{\text{0}}^{\text{4}}}$, the proposed scheme tends to converge, 
but owing to the dynamic nature of the environment and the system's inherent uncertainty, the convergence values of the proposed algorithm fluctuates within a certain range.
Compared to the other two algorithms, the proposed approach exhibits a lower convergence value.
That is because the REDQ algorithm is based on the maximum entropy framework, 
which encourages agent to explore more behavioral strategies rather than prematurely focusing on a specific strategy. 
This helps to avoid getting stuck in local optima and increases the probability of discovering the global optimal strategy.
\begin{figure}[t]
  \centering
  \includegraphics[width=7.7cm]{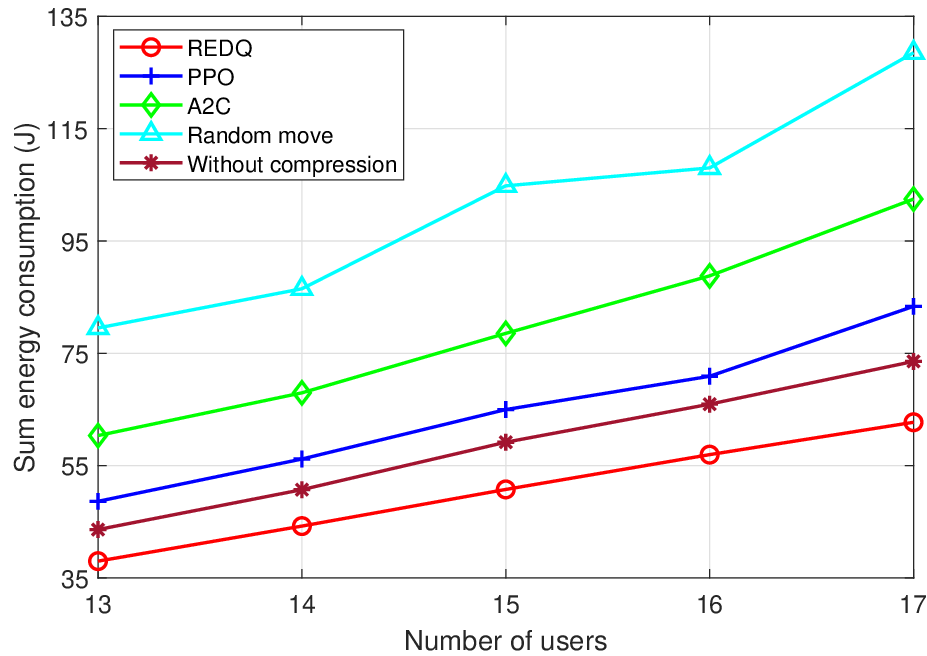}
  \caption{Performance comparison versus different number of users.}
  \label{fig:user}
\end{figure}

Fig. \ref{fig:user} illustrates the total energy consumption across varying numbers of users.
We add a comparison scheme called \textit{without compression} to highlight the effectiveness of data compression technology.
It is evident that as the number of users becomes large, energy consumption values for all schemes accordingly increases. 
The proposed scheme has the best performance, while 
the \textit{random move} scheme incurs the highest energy cost, 
which indicates that trajectory planning plays a crucial role in determining system performance.
As the user count rises, the gap between the proposed solution and \textit{without compression} scheme becomes more apparent, 
which means that data compression can effectively reduce user energy consumption.
In addition, compared with PPO and A2C algorithms, the proposed scheme reduces energy consumption by about $24.8\%$ and $38.8\%$, respectively, 
at a simulated user count of $17$, 
which reflects the proposed scheme offers a more efficient method for trajectory planning and resource allocation.

\begin{figure}[t]
  \centering
  \includegraphics[width=7.7cm]{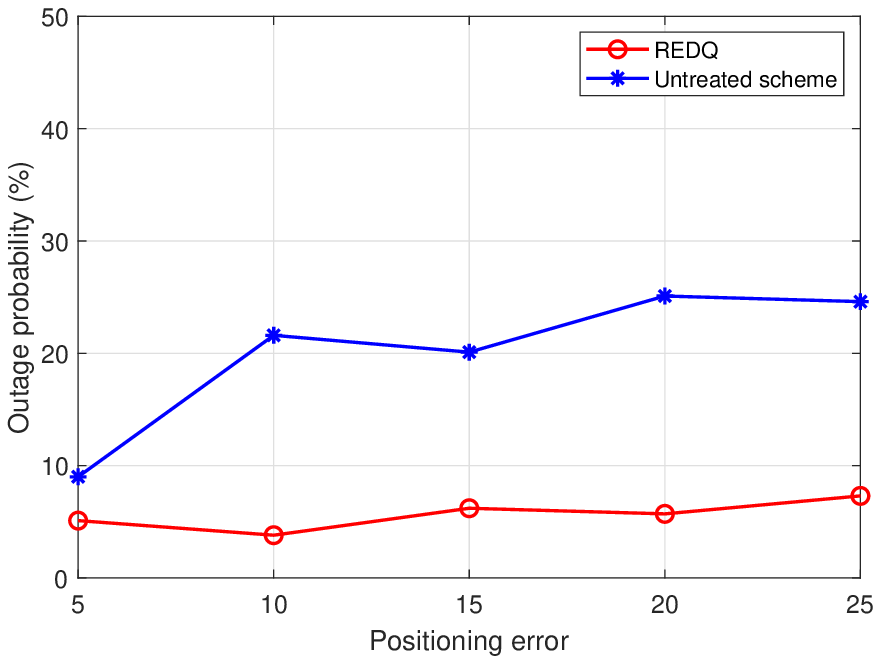}
  \caption{Outage probability under different positioning errors.}
  \label{fig:out}
\end{figure}

\begin{figure}[t]
  \centering
  \includegraphics[width=7.7cm]{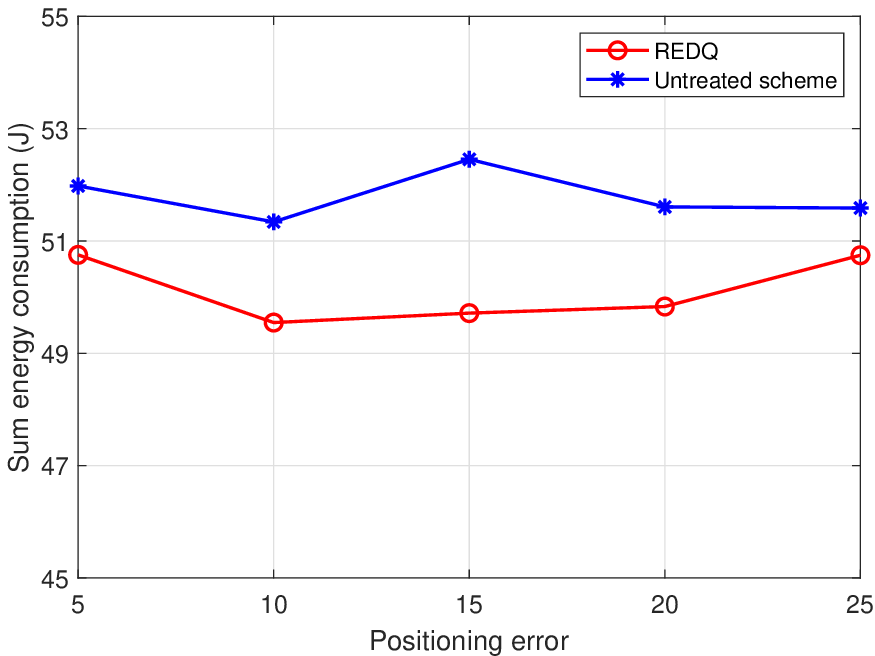}
  \caption{Energy consumption under different positioning errors.}
  \label{fig:outE}
\end{figure}

Fig. \ref{fig:out} and Fig. \ref{fig:outE} respectively show the outage probability and energy consumption under different position errors,
where the outage probability represents the probability of the UAV violating speed constraint. 
From Fig. \ref{fig:out}, it can be observed that as the position error increases, 
the outage probability of the proposed scheme can always remain below $10\%$ and fluctuate steadily,
while the outage probability of the untreated scheme is much higher than that of the proposed scheme and shows an overall upward trend.
This means that the proposed scheme can effectively control the flight of the UAV.
When combined with Fig. \ref{fig:outE}, 
it is clear that the energy consumption of the proposed scheme is consistently lower than that of the untreated scheme. 
This reflects that the UAV jittering has a certain impact on system performance, 
and the proposed solution in this paper can effectively alleviate this problem.

\begin{figure}[t]
  \centering
  \includegraphics[width=7.7cm]{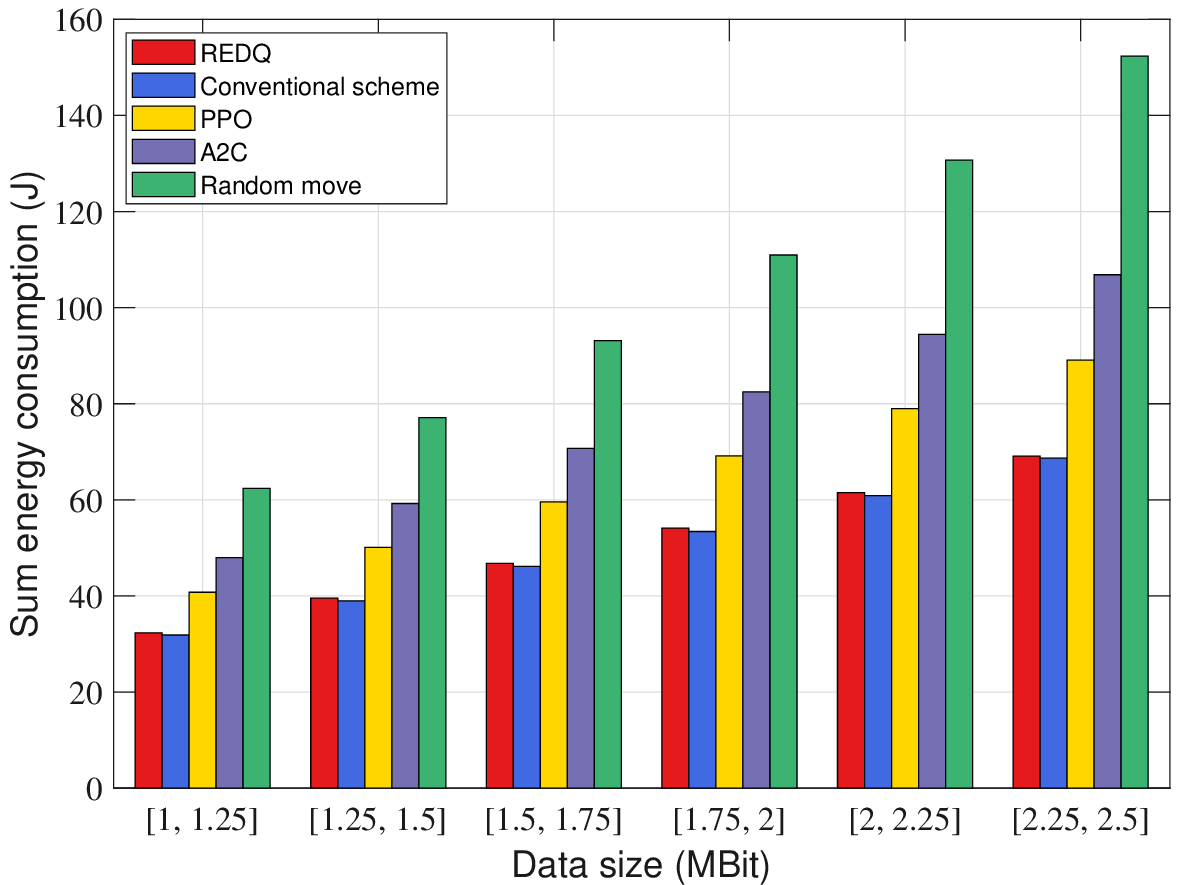}
  \caption{Performance comparison under different task data sizes.}
  \label{fig:task}
\end{figure}

Fig. \ref{fig:task} evaluates the performance obtained by the five schemes versus different data size of task.
The scheme called \textit{conventional scheme} is added for comparison, which do not take into account the UAV jittering. 
The bar chart reveals that as the task data volume grows, the energy costs for all schemes increase.
Compared to PPO, A2C, and \textit{random move}, the proposed scheme has reduced energy consumption by approximately $21.5\%$, $33.9\%$, and $49.8\%$, 
respectively, with the setting of task size of [$1.5$, $1.75$] $\mathrm{Mbits}$,
which further highlights the advantages of the proposed solution.
Moreover, 
the \textit{conventional scheme} has a slightly lower energy cost compared to the proposed solution, 
because the \textit{conventional scheme} can more accurately control the UAV flight and provide better trajectory planning.

\begin{figure}[t]
  \centering
  \includegraphics[width=7.6cm]{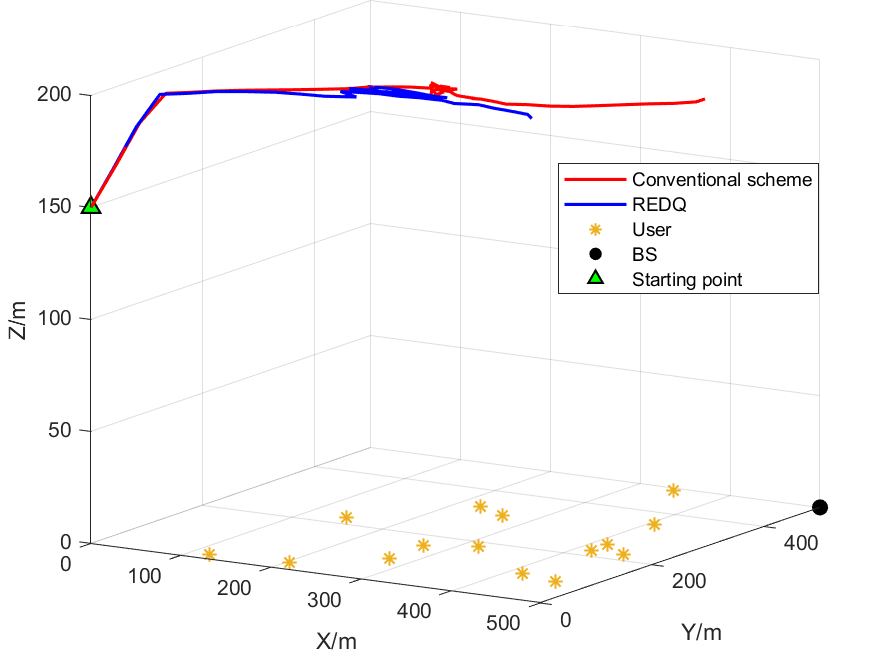}
  \caption{Example of 3D trajectory of the UAV.}
  \label{fig:UAV3DGT}
\end{figure}
\begin{figure}[t]
  \centering
  \includegraphics[width=7.6cm]{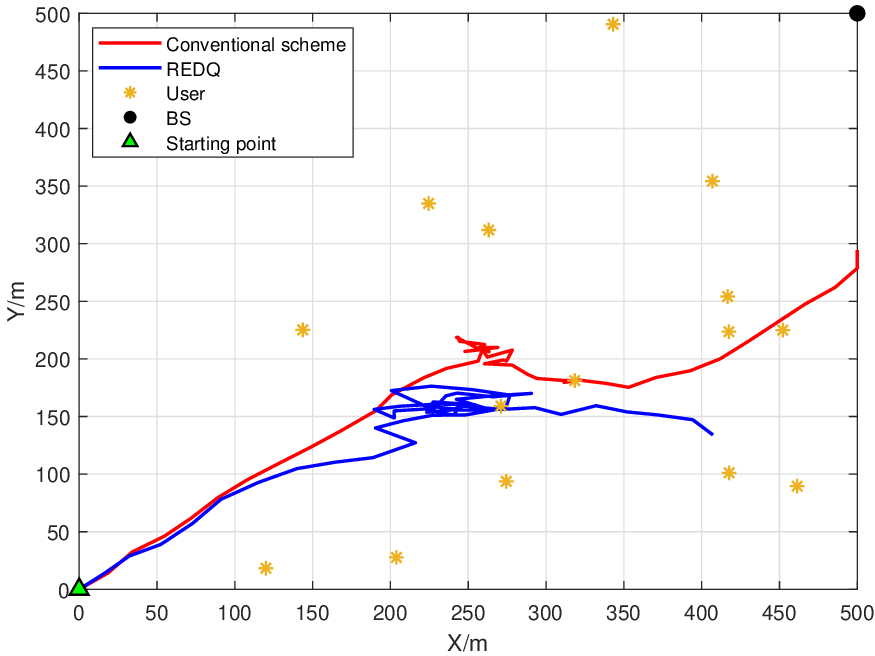}
  \caption{Example of 2D trajectory of the UAV.}
  \label{fig:UAV2DGT}
\end{figure}
Fig. \ref{fig:UAV3DGT} and Fig. \ref{fig:UAV2DGT} illustrate the 2D and 3D trajectories of the UAV, respectively.
By combining the two images, it can be observed that the UAV increases its flight altitude 
while flying towards the central area where terminal users are distributed.
The UAV attempts to maintain a close distance with all users. 
The reason is that being close to the users can reduce path loss, 
and increasing the flight altitude appropriately can enlarge the elevation angle from the user to the UAV, 
thereby obtaining greater effective fading power, which can achieve higher transmission rates, and reduce users' offloading energy consumption.
The difference in trajectory between the two schemes is that the \textit{conventional scheme} is more flexible and can actively explore more areas, 
while the proposed scheme is more conservative,
as shown in the figure where the trajectory of the proposed scheme spends more time staying in a certain area.
The reason may be that the proposed scheme need to consider uncertain factors, which results in the loss of some degrees of freedom.

\section{CONCLUSION}
In this paper, we considered the trajectory uncertainty caused by UAV jittering in 3D space and proposed an energy optimization scheme based on data compression technology. 
The scheme aimed to reduce the energy consumption of terminal users while ensuring UAV trajectory robustness. 
Given the non-convexity of the formulated problem and the dynamic characteristics of the scenario, 
we applied a DRL algorithm to solve this complex optimization problem. 
Simulation results demonstrated that the proposed scheme outperforms the benchmarks.
In future work, we will further study the multi-UAV-assisted MEC scenario for providing a wider coverage and more flexible computing services, helping to further improve system performance.

\bibliographystyle{IEEEtran}
\bibliography{IEEEabrv,refs}

\begin{thebibliography}{10}
\providecommand{\url}[1]{#1}
\csname url@samestyle\endcsname
\providecommand{\newblock}{\relax}
\providecommand{\bibinfo}[2]{#2}
\providecommand{\BIBentrySTDinterwordspacing}{\spaceskip=0pt\relax}
\providecommand{\BIBentryALTinterwordstretchfactor}{4}
\providecommand{\BIBentryALTinterwordspacing}{\spaceskip=\fontdimen2\font plus
\BIBentryALTinterwordstretchfactor\fontdimen3\font minus
  \fontdimen4\font\relax}
\providecommand{\BIBforeignlanguage}[2]{{%
\expandafter\ifx\csname l@#1\endcsname\relax
\typeout{** WARNING: IEEEtran.bst: No hyphenation pattern has been}%
\typeout{** loaded for the language `#1'. Using the pattern for}%
\typeout{** the default language instead.}%
\else
\language=\csname l@#1\endcsname
\fi
#2}}
\providecommand{\BIBdecl}{\relax}
\BIBdecl

\bibitem{UAVIntro}
X.~Diao, W.~Yang, L.~Yang, and Y.~Cai, ``{UAV}-relaying-assisted multi-access
  edge computing with multi-antenna base station: Offloading and scheduling
  optimization,'' \emph{IEEE Transactions on Vehicular Technology}, vol.~70,
  no.~9, pp. 9495--9509, Sep. 2021.

\bibitem{UAVIntroE1}
A.~Mondal, D.~Mishra, G.~Prasad, and A.~Hossain, ``Joint optimization framework
  for minimization of device energy consumption in transmission rate
  constrained {UAV}-assisted {IoT} network,'' \emph{IEEE Internet of Things
  Journal}, vol.~9, no.~12, pp. 9591--9607, Jun. 2022.

\bibitem{UAVIntroE2}
C.~Zhan, H.~Hu, X.~Sui, Z.~Liu, and D.~Niyato, ``Completion time and energy
  optimization in the {UAV}-enabled mobile-edge computing system,'' \emph{IEEE
  Internet of Things Journal}, vol.~7, no.~8, pp. 7808--7822, Aug. 2020.

\bibitem{UAVIntroE3}
B.~Xu, Z.~Kuang, J.~Gao, L.~Zhao, and C.~Wu, ``Joint offloading decision and
  trajectory design for {UAV}-enabled edge computing with task dependency,''
  \emph{IEEE Transactions on Wireless Communications}, vol.~22, no.~8, pp.
  5043--5055, Aug. 2023.

\bibitem{UAVIntroE4}
H.~Hao, C.~Xu, W.~Zhang, S.~Yang, and G.-M. Muntean, ``Joint task offloading,
  resource allocation, and trajectory design for multi-{UAV} cooperative edge
  computing with task priority,'' \emph{IEEE Transactions on Mobile Computing},
  vol.~23, no.~9, pp. 8649--8663, Sep. 2024.

\bibitem{DRL2}
R.~Zhang, H.~Du, D.~Niyato, J.~Kang, Z.~Xiong, A.~Jamalipour, P.~Zhang, and
  D.~I. Kim, ``Generative {AI} for space-air-ground integrated networks,''
  \emph{IEEE Wireless Communications}, vol.~31, no.~6, pp. 10--20, Dec. 2024.

\bibitem{compressionintro}
Y.~Zeng and D.~Liu, ``Computation offloading based on improved sparrow search
  algorithm in edge computing scenario,'' in \emph{Proc. International
  Conference on Computing, Communication, Perception and Quantum Technology
  (CCPQT)}, Xiamen, China, 2022, pp. 231--236.

\bibitem{losslessCompress}
D.~Xu, Q.~Li, and H.~Zhu, ``Energy-saving computation offloading by joint data
  compression and resource allocation for mobile-edge computing,'' \emph{IEEE
  Communications Letters}, vol.~23, no.~4, pp. 704--707, Apr. 2019.

\bibitem{lossyCompress}
M.~Hosseinzadeh, N.~Hudson, X.~Zhao, H.~Khamfroush, and D.~E. Lucani, ``Joint
  compression and offloading decisions for deep learning services in 3-tier
  edge systems,'' in \emph{Proc. IEEE International Symposium on Dynamic
  Spectrum Access Networks (DySPAN)}, Los Angeles, USA, 2021, pp. 254--261.

\bibitem{lossylossless}
S.~Lu, Q.~Xia, X.~Tang, X.~Zhang, Y.~Lu, and J.~She, ``A reliable data
  compression scheme in sensor-cloud systems based on edge computing,''
  \emph{IEEE Access}, vol.~9, pp. 49\,007--49\,015, 2021.

\bibitem{Compress3Loc}
J.~Ren, G.~Yu, Y.~Cai, and Y.~He, ``Latency optimization for resource
  allocation in mobile-edge computation offloading,'' \emph{IEEE Transactions
  on Wireless Communications}, vol.~17, no.~8, pp. 5506--5519, Aug. 2018.

\bibitem{CompressRatio}
J.-B. Wang, J.~Zhang, C.~Ding, H.~Zhang, M.~Lin, and J.~Wang, ``Joint
  optimization of transmission bandwidth allocation and data compression for
  mobile-edge computing systems,'' \emph{IEEE Communications Letters}, vol.~24,
  no.~10, pp. 2245--2249, Oct. 2020.

\bibitem{UAVCompress}
K.~Cheng, X.~Fang, and X.~Wang, ``Energy efficient edge computing and data
  compression collaboration scheme for {UAV}-assisted network,'' \emph{IEEE
  Transactions on Vehicular Technology}, vol.~72, no.~12, pp. 16\,395--16\,408,
  Dec. 2023.

\bibitem{UAVjitterintro}
M.~Zhan and S.~Xu, ``Delay-aware distributionally robust trajectory
  {UAV}-assisted {MEC} with uncertain task size,'' in \emph{Proc. IEEE Wireless
  Communications and Networking Conference (WCNC)}, Dubai, United Arab
  Emirates, 2024, pp. 1--6.

\bibitem{robustJitter}
H.~Wu, Y.~Wen, J.~Zhang, Z.~Wei, N.~Zhang, and X.~Tao, ``Energy-efficient and
  secure air-to-ground communication with jittering {UAV},'' \emph{IEEE
  Transactions on Vehicular Technology}, vol.~69, no.~4, pp. 3954--3967, Apr.
  2020.

\bibitem{robustJitter2}
X.~Tang, H.~Zhang, R.~Zhang, D.~Zhou, Y.~Zhang, and Z.~Han, ``Robust trajectory
  and offloading for energy-efficient {UAV} edge computing in industrial
  internet of things,'' \emph{IEEE Transactions on Industrial Informatics},
  vol.~20, no.~1, pp. 38--49, Jan. 2024.

\bibitem{robustJitter3}
W.~Lee and K.~Lee, ``Robust trajectory and resource allocation for {UAV}
  communications in uncertain environments with no-fly zone: A deep learning
  approach,'' \emph{IEEE Transactions on Intelligent Transportation Systems},
  vol.~25, no.~10, pp. 14\,233--14\,244, Oct. 2024.

\bibitem{UAVRelayChannel}
Q.~Qi, T.~Shi, K.~Qin, and G.~Luo, ``Completion time optimization in
  {UAV}-relaying-assisted {MEC} networks with moving users,'' \emph{IEEE
  Transactions on Consumer Electronics}, vol.~70, no.~1, pp. 1246--1258, Feb.
  2024.

\bibitem{UAVRelayChannel2}
S.~Sun, G.~Zhang, H.~Mei, K.~Wang, and K.~Yang, ``Optimizing multi-{UAV}
  deployment in 3-{D} space to minimize task completion time in {UAV}-enabled
  mobile edge computing systems,'' \emph{IEEE Communications Letters}, vol.~25,
  no.~2, pp. 579--583, Feb. 2021.

\bibitem{compressionEqua}
X.~Li, C.~You, S.~Andreev, Y.~Gong, and K.~Huang, ``Wirelessly powered crowd
  sensing: Joint power transfer, sensing, compression, and transmission,''
  \emph{IEEE Journal on Selected Areas in Communications}, vol.~37, no.~2, pp.
  391--406, Feb. 2019.

\bibitem{UAVEnergy}
N.~Lin, H.~Tang, L.~Zhao, S.~Wan, A.~Hawbani, and M.~Guizani, ``A {PDDQNLP}
  algorithm for energy efficient computation offloading in {UAV}-assisted
  {MEC},'' \emph{IEEE Transactions on Wireless Communications}, vol.~22,
  no.~12, pp. 8876--8890, Dec. 2023.

\bibitem{DRL1}
R.~Zhang, H.~Du, Y.~Liu, D.~Niyato, J.~Kang, S.~Sun, X.~Shen, and H.~V. Poor,
  ``Interactive {AI} with retrieval-augmented generation for next generation
  networking,'' \emph{IEEE Network}, vol.~38, no.~6, pp. 414--424, Nov. 2024.

\bibitem{redq}
\BIBentryALTinterwordspacing
X.~Chen, C.~Wang, Z.~Zhou, and K.~W. Ross, ``Randomized ensembled double
  {Q}-learning: Learning fast without a model,'' \emph{arXiv preprint
  arXiv:2101.05982}, 2021. [Online]. Available:
  \url{https://arxiv.org/abs/2101.05982.}
\BIBentrySTDinterwordspacing

\end{thebibliography}

\end{document}